# Simulation-based optimization of a multilayer $^{10}$B-RPC thermal neutron detector

A. Morozov[1,*], L.M.S. Margato[1] and I. Stefanescu[2]

[1] LIP-Coimbra, Department of Physics, Coimbra University, 3004-516, Coimbra, Portugal

[2] European Spallation Source ERIC (ESS), P.O. Box 176, SE-22100, Lund, Sweden

* E-mail: andrei@coimbra.lip.pt

## Abstract:

A Monte Carlo simulation-based optimization of a multilayer $^{10}$B-RPC thermal neutron detector is performed targeting an increase in the counting rate capability while maintaining high (>50%) detection efficiency for thermal neutrons. The converter layer thicknesses of individual RPCs are optimized for several configurations of a detector containing a stack of 10 double gap RPCs. The results suggest that it is possible to reach a counting rate which is by a factor of eight higher in comparison to the rate of a detector with only one double-gap RPC. The effect of neutron scattering inside the detector contributing to the background is analyzed and design modifications of the first detector prototype, tested at neutron beam, are suggested.

# 1  Introduction

Recently we have introduced a concept of a $^{10}$B-RPC position sensitive thermal neutron detector [1] based on resistive plate chambers (RPCs, [2]) lined with $^{10}$B$_4$C neutron converters (see, e.g., [3]) in a multilayer configuration and then tested feasibility of this concept experimentally at neutron beam [4]. We have demonstrated that a $^{10}$B-RPC detector with 20 converter layers of $^{10}$B$_4$C has a potential to reach detection efficiency of 50% - 70% for thermal neutrons, sub-millimeter (down to 0.1 mm) spatial resolution, and sub-nanosecond time resolution [1,4].

One of the challenges for detectors based on the RPC technology is their operation at high counting rates: for RPCs with electrodes made of a material with high electrical resistivity, such as, for example, widely used soda-lime glass (~$10^{13}$ Ω·cm), the maximum counting rate density is typically below 1 kHz per cm$^2$. This limitation originates from the fact that with increase of the particle flux the current flowing through the resistive electrode increases and a drop of potential develops at the electrode surface facing the gas gap. The resulting decrease of the potential across the gas gap reduces the gas amplification leading to a decrease in the average induced charge per event. Thus, with increase of the counting rate density a larger fraction of events is discriminated and the detection efficiency decreases. Targeting increase in the maximum counting rate density, operation of RPCs with electrodes made of materials with low resistivity is currently being investigated. Counting rates of tens of kHz per cm$^2$ have already been reported for RPCs with resistive electrodes made of ceramics [5,6] and several types of glasses [7-9].

Operation of a single-gap $^{10}$B-RPC detector prototype with counting rate densities up to 1 kHz/cm$^2$ without loss of efficiency has been reported in our previous study [4]. One of the approaches foreseen to improve the counting rate capability of the detector is to increase the number of RPCs [1] distributing the detection of neutrons impinging the detector in the same area over multiple RPCs. Recently we have tested a detector prototype with a stack of 20 RPCs, oriented normally to the beam, each having the same thickness of the $^{10}$B$_4$C converter layer [10]. The thickness was optimized to maximize the overall detection efficiency. However, the maximum counting rate of such detector does not scale linearly with the number of RPCs in the stack. The neutron flux decreases exponentially along the RPC stack mainly due to attenuation of the beam in the converter layers passed by the beam. Since the detection efficiency of all RPCs is the same, the counting rate density of the individual RPCs also reduces exponentially. Therefore, in order to equalize as much as possible the counting rates of all RPCs in the stack, the thicknesses of individual converter layers have to be optimized.

Another aspect which has to be evaluated is the fraction of detection events originating from the neutrons which had at least one elastic scattering before detection. These events do not provide useful spatial information and only contribute to the background. The materials having the strongest contribution to the scattering have to be identified and realistic modifications of the detector design minimizing scattering have to be considered.

In this paper we continue the simulation study reported in [1] by introducing elastic scattering of neutrons. We cross-compare the simulation results on the energy deposition in the gas gaps obtained with several simulation toolkits (Geant4 [11], ANTS2 [12], ANTS2 + NCrystal [13]). We also compare the estimated detection efficiencies obtained using these toolkits with the results of the measurements [10] performed with a $^{10}$B-RPC detector prototype with 20 converter layers at the TREFF [14] neutron beam line (3.6 meV, $\lambda$ = 4.73 Å) at the Research Neutron Source Heinz Maier-Leibnitz (FRM II). We present an analysis of the contributions to the neutron scattering and absorption given by the detector materials and define realistic changes in the detector design which can reduce neutron scattering inside the detector. Finally, we present an optimization study in which we adjust the thicknesses of the converter layers of individual RPCs aiming to increase the maximum counting rate capability while keeping the detection efficiency as high as possible.

## 2 Methods

### 2.1 Simulation tools

Two simulation toolkits are used in this study: Geant4 [11] version 10.5.1 and ANTS2 [12] version 4.21. Simulations with Geant4 are performed with the QGSP_BIC_HP and, alternatively, with the QGSP_BIC_AllHP reference physics lists. In this version of Geant4, QGSP_BIC_HP activates by default high precision models for electromagnetic processes (option 4). Both physics lists use a high precision model for transport of thermal neutrons. The production cuts are set to 0.01 mm. The maximum tracking step is limited to 0.01 mm in the RPC gas gaps and to 0.01 µm in the $^{10}$B$_4$C neutron converters.

The ANTS2 toolkit offers an infrastructure for semi-supervised optimization of detector geometry. Since it is a custom software package and a number of significant changes have been introduced after publication of the article describing the toolkit [12], a more detailed description of the relevant features is presented below.

ANTS2 features a simple neutron transport model, optimized for simulations of thermal neutron detectors. Two simulation options are available. Using the first one, all materials are assumed to be gas mixtures of the corresponding isotopes at the temperature of 300 K. Isotope velocities are sampled directly from the Maxwell distribution. Elastic scattering is considered without coherent effects (isotropic in the center of mass frame of the atom – neutron pair). The total elastic (N,EL) and total non-elastic (N,NON) cross-sections versus energy are taken from the ENDF/B-VII.1 database [15], or from JEFF-3.2 [16] and JENDL-4.0u2 [17] databases if the data are missing in ENDF/B for a particular isotope. All cross-section data are downloaded using the IAEA online data services [18]. For the second simulation option the only difference is that simulation of elastic scattering in metals is delegated to NCrystal library [13] which takes into account coherent effects.

After a neutron is absorbed by a boron-10 nucleus, the following particles are generated: in 93.9% of the cases it is an alpha particle (1.47 MeV) plus a lithium-7 ion (0.84 MeV) plus a gamma ray (0.478 MeV), and in 6.1% of the cases it is an alpha particle (1.78 MeV) plus a lithium-7 ion (1.01 MeV). The interaction of alpha particles and lithium-7 ions with the detector materials is simulated using the stopping power provided by SRIM [19]. The interaction cross-sections for gamma rays are taken from the XCOM database [20]. For gamma rays, only photoelectric effect, Compton scattering and pair production are simulated in ANTS2, and the fluorescence and Auger electron emission are not taken into account.

## 2.2 Detector model

A detailed description of the $^{10}$B-RPC detector concept can be found in [1], so only a brief description of the detector is given here. The design and the parameter values described in this section are the ones of the first experimental prototype [10] tested at FRM II.

The detector is comprised of 10 double-gap hybrid RPCs. Each of them features an aluminium cathode (90 x 90 x 0.5 mm$^3$), covered on both sides with a 1.15 µm thick layer of $^{10}$B$_4$C neutron converter, enriched to 97% in boron-10. The converters are separated from the soda-lime glass anodes (100 x 100 x 0.5 mm$^3$) by 0.35 mm wide gas gaps. The detector is filled with tetrafluoroethane (C$_2$F$_4$H$_2$) at atmospheric pressure. The neighboring double-gap RPCs share a multilayer printed circuit board (PCB) with pick-up electrodes for XY position readout [1]. Each PCB has three planes of copper electrodes (0.018 mm thick), separated by 0.025 mm polyimide films (see figure 1 and 2). The distance between the neighboring RPCs and PCBs is 0.2 mm.

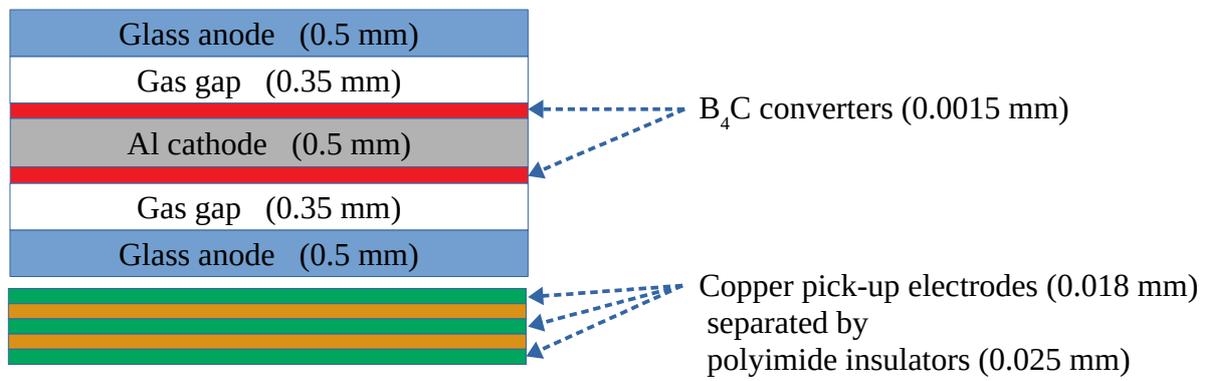

Figure 1. Schematic drawing (not to scale) of an elementary block of a multilayer $^{10}$B-RPC detector.

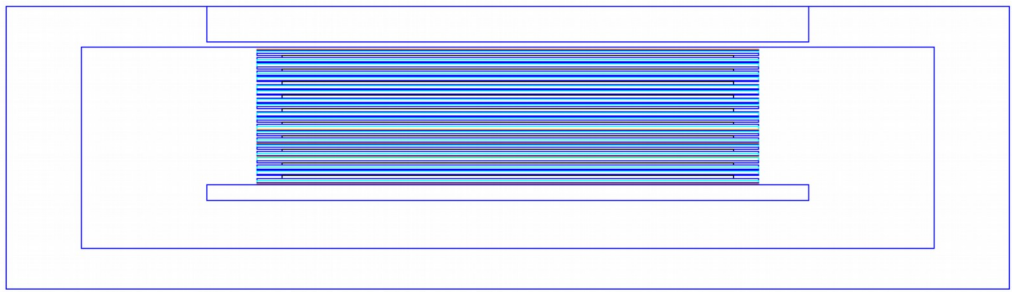

Figure 2. Side view of the ANTS2 model of the detector featuring a stack of 10 double-gap hybrid RPCs. The stack (27.2 mm total width) consists of 10 elementary blocks presented in figure 1. The detector enclosure (aluminium box) has a 1 mm thick entrance window (figure top) and a 3 mm thick aluminium mounting plate, situated under the RPC stack.

The RPC stack (27.2 mm total width) is mounted inside an aluminium enclosure with a 1 mm thick entrance window. The mounting plate (3 mm thick aluminium) is situated on the opposite side from the window (see figure 2).

In the simulations, a molar composition $SiO_2$:72.98 + $Na_2O$:14 + $CaO$:7 + $MgO$:4 + $Al_2O_3$:2 + $K_2O$:0.02 and a density of 2.53 g/cm$^3$ are used for the RPC anodes (typical for soda-lime glasses). The $B_4C$ molar composition, provided by the manufacturer (ESS Detector Coatings Workshop), is B:81.7 + C:17 + H:0.7 + O:0.4 + N:0.2 and the density is 2.242 g/cm$^3$. The polyimide ($C_{41}H_{22}N_4O_{11}$) density is 1.42 g/cm$^3$. We assume that the cathodes and the detector enclosure are made of pure aluminium, and that the signal pick-up electrodes are made of pure copper.

## 2.3 Semi-automatic multi-parameter optimization in ANTS2

Simulation-based detector optimization can be very time consuming if it has to be performed over several parameters which have an "entangled" effect on the performance of the detector. In such conditions it is not possible to perform optimization independently over each parameter, and, typically, a large number of simulations covering a grid in the multi-dimensional parameter space has to be conducted.

An alternative approach, applied in this study, is to implement a minimization algorithm operating with a custom cost function. On each call from the minimizer, the function receives a value for each parameter, modifies according to them the detector model, starts a simulation of a certain number of neutron events, processes the results and, finally, calculates the value of a user-defined parameter (the cost value) which is then returned to the minimizer.

An infrastructure to implement this approach is available in ANTS2: it is possible, using the scripting system of the toolkit, to define such a minimization function. The function has access to the methods allowing to modify the detector model, configure and execute simulations, and retrieve the simulation results. The minimizer, also accessible from script, is configured to run the simplex algorithm implemented in the CERN ROOT library [21].

# 3 Results and discussion

## 3.1 Detection efficiency

To estimate the detection efficiency of the $^{10}$B-RPC detector, simulations are performed in which $10^6$ neutrons are generated in a pencil beam passing the geometric center of the entrance window and oriented normally to it. The neutron energy is 3.656 meV, which corresponds to the energy of the TREFF beam line at FRM II where the tests of the first detector prototype have been performed [10].

Distributions of the energy deposited in the gas gaps per neutron capture event, simulated with Geant4 and ANTS2 are shown in figure 3 (left). The distributions obtained with Geant4 using the QGSP_BIC_HP and QGSP_BIC_AllHP physics list are practically identical. Also, both profiles obtained with ANTS2 (directly or using the NCrystal library for simulation of scattering in metals) are essentially the same. There are only two minor features which are different in the results of the two toolkits. The first one is a slight shift in the main peak at ≈ 0.3 MeV. The second one is a more frequent deposition of low energies (especially below 10 keV) predicted by Geant4, a fact which is expected since Geant4 generates and traces secondary gamma rays and electrons, while ANTS2 does not.

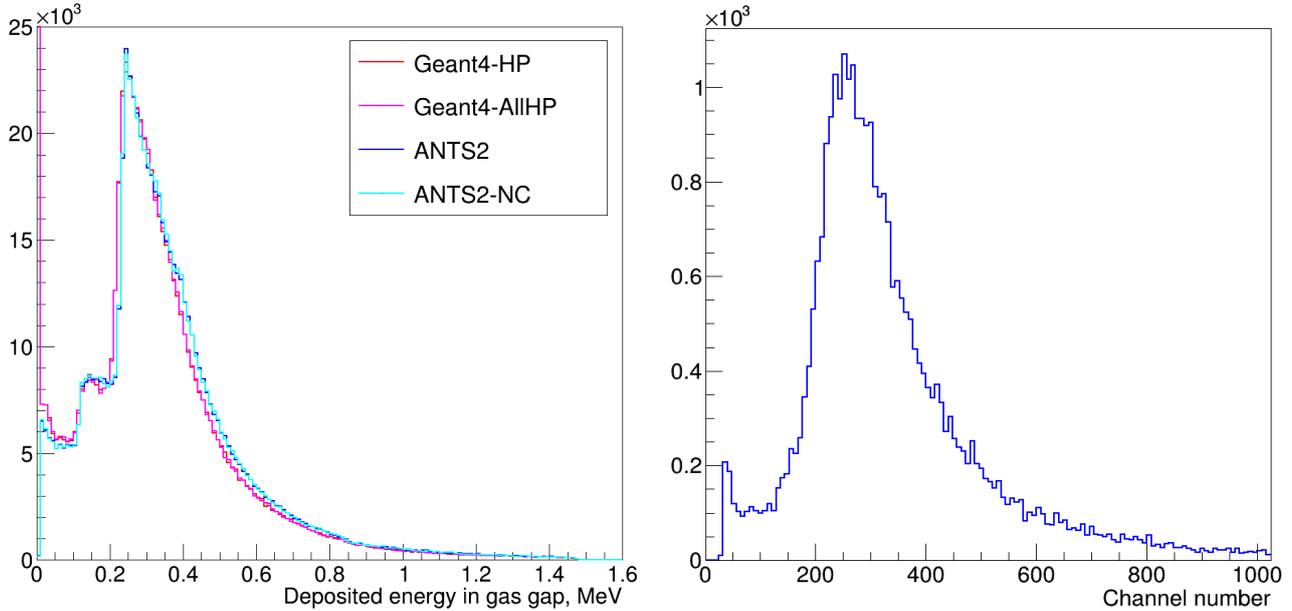

Figure 3. Left hand side: Distribution of the energy deposited in the gas gaps per neutron event simulated using Geant4 (QGSP_BIC_HP and QGSP_BIC_AllHP physics lists) and ANTS2 (directly or using the NCrystal library for simulation of scattering in metals). Right hand side: Experimental pulse height spectrum of the charge induced in the cathodes which was recorded with the detector prototype irradiated by neutron beam (λ = 4.73 Å).

Figure 3 (right) shows a pulse height spectrum of the charge induced in the cathodes which was recorded with the detector prototype irradiated by a neutron beam (λ = 4.73 Å) [10]. The simulated and experimental profiles show a high degree of similarity. However, the distributions of the deposited and the induced charge should be compared only on a qualitative level: note that for events with the same deposited energy the induced charge can vary depending on the spatial distribution of the energy deposition and development of the electron avalanche(s) inside the gas gap.

In this study we consider that a neutron capture event is detected if it results in deposition in a gas gap an amount of energy above a certain threshold [1]. Setting this threshold to 100 keV, we obtain the following total detection efficiencies: 59.0%, 59.0%, 60.3% and 60.3% for Geant4-HP, Geant4-AllHP, ANTS2 and ANTS2-NC, respectively (statistical uncertainties are ±1 in the last digit). The contributions to the total detection efficiency from the individual double gap RPCs are shown in figure 4.

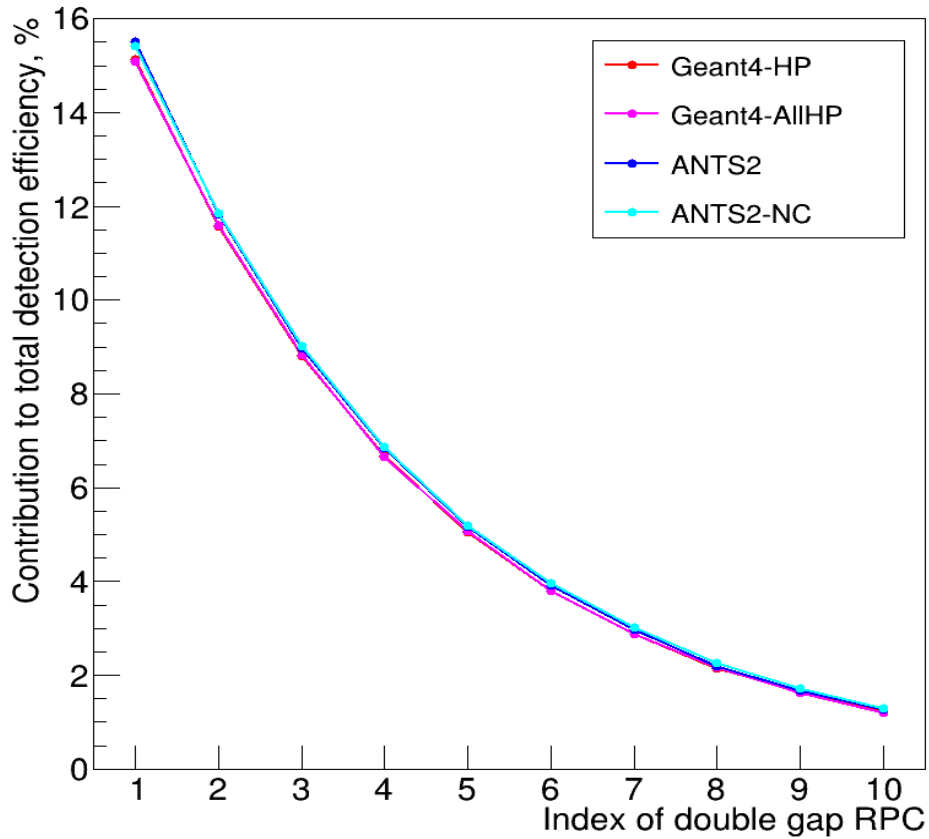

Figure 4. Contributions to the total detection efficiency from the individual double gap RPCs. Index of 1 is assigned to the RPC in the stack closest to the entrance window. The total detection efficiency is the sum of the values shown for the individual double gap RPCs. The detection threshold energy is 100 keV. The statistical uncertainties are smaller than the size of the markers.

Figure 5 shows the contributions to the total detection efficiency from the individual double gap RPCs obtained in ANTS2 simulations for two threshold energies (25 and 100 keV) as well as the experimental data recorded with the detector prototype [10]. The total detection efficiency measured with the prototype is (62.1 ± 4.5)%, which is in a good agreement with the values obtained in the simulations: 64.3% and 60.3% for the thresholds of 25 keV and 100 keV, respectively (statistical uncertainties are ± 1 in the last digit).

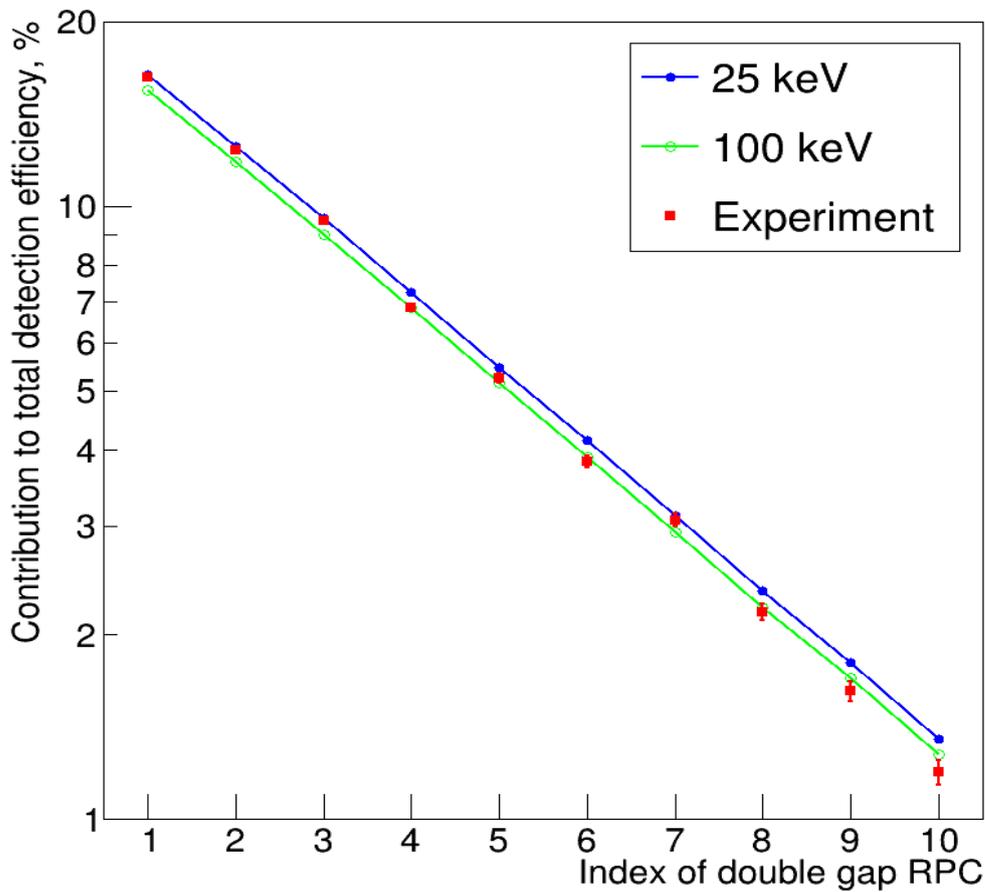

Figure 5. Contributions to the total detection efficiency calculated based on an ANTS2 simulation for the threshold energies of 25 keV (filled circles) and 100 keV (open circles) as well as the experimental data (squares) obtained with the first detector prototype at neutron beam. For experimental data the error bars show the statistical uncertainties and the systematic uncertainty is of a factor of 0.07 of the contribution value. For simulation data the statistical uncertainties are smaller than the size of the markers.

## 3.2 Effect of the detector materials: neutron absorption and scattering

Table 1 shows the fractions (in percents) of the number of neutrons removed from the beam in the materials of the stack of 10 double-gap RPCs due to scattering and absorption. The data are simulated for a monochromatic (3.656 meV) beam with $10^6$ neutrons hitting the detector normally at the center of the first RPC electrode. The effect of the detector enclosure (the aluminium box with the neutron entrance window) and the mounting plate is excluded from the data.

|  | ANTS2 | | ANTS2 + NC | | Geant4 (HP) | | Geant4 (AllHP) | |
|---|---|---|---|---|---|---|---|---|
|  | Scat | Abs | Scat | Abs | Scat | Abs | Scat | Abs |
| Glass, 10 mm | 9.22 | 0.65 | 9.44 | 0.67 | 9.17 | 0.67 | 9.15 | 0.65 |
| Al, 5 mm | 1.56 | 0.59 | 0.14 | 0.59 | 1.50 | 0.59 | 1.52 | 0.58 |
| Polyimide, 0.6 mm | 3.71 | * | 3.80 | * | 3.57 | * | 3.56 | * |
| Cu, 0.6 mm | 1.40 | 1.67 | 0.18 | 1.70 | 1.37 | 1.72 | 1.37 | 1.69 |
| $B_4C$, 0.03 mm | * | 76.0 | * | 77.8 | * | 76.3 | * | 76.3 |
| Gas, 11 mm | 0.17 | * | 0.17 | * | 0.17 | * | 0.17 | * |

Table 1. Fractions (in percents) of the number of neutrons removed from the beam in the materials of the stack of 10 double-gap hybrid RPCs due to scattering (Scat) and absorption (Abs). The values less than 0.1% are indicated with a * symbol. The combined thickness of each material is shown in the first column. The statistical uncertainty is less than ±5 in the last digit.

The results obtained with ANTS2 and Geant4 are quite similar. The only significant difference is a much weaker scattering in both metals given by the simulations performed in ANTS2 using the NCrystal library. This fact can be explained by taking into account that NCrystal computes the scattering cross-section using information on the structure of the crystal lattice, thus evaluating both the coherent and incoherent scattering effects, while the standalone ANTS2 and Geant4 assume that the scattering cross-section is the sum of the incoherent and the coherent components for all neutron energies (see figure 6).

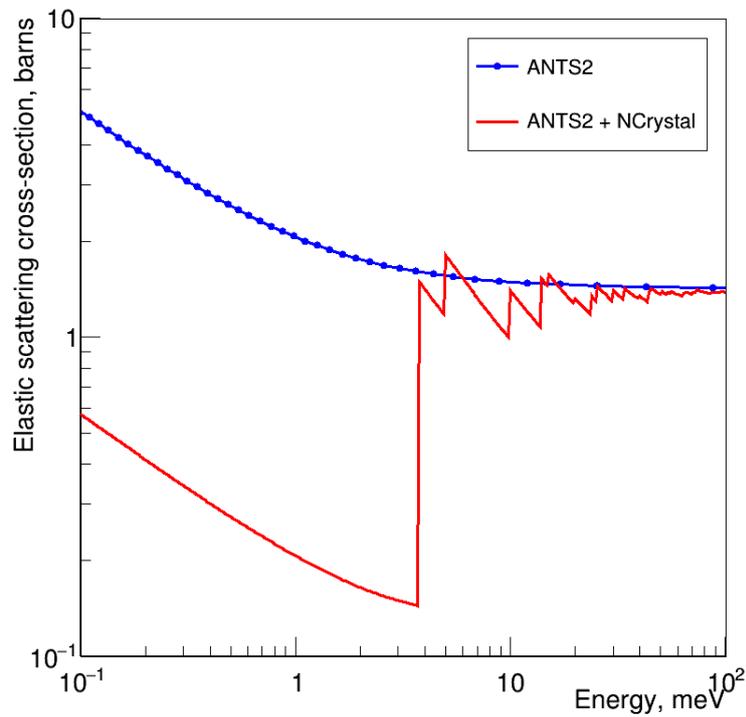

Figure 6. Scattering cross-section vs energy used in ANTS2 (line with dots) and ANTS2+NCrystal (straight line) simulations. The first increase in the cross-section used in ANTS+NCrystal simulations is at an energy higher than the energy of neutrons in the beam (3.656 meV).

Note that scattering in aluminium and copper simulated with NCrystal is performed assuming pure metals. Since the detector components are manufactured from alloys, their lattice structure has to be determined and introduced in the toolkit in order to perform more realistic simulations. However, the results presented in figure 3 and 4 as well as the relatively small scattering fractions for these materials (see table 1) suggest that for the detector design considered here this step is not mandatory.

ANTS2 simulations predict a ratio of 0.133 for the number of the detected indirect events (at least one elastic scattering before neutron capture) to the total number of the detected events. The indirect events contribute to the detector background, therefore, modifications of the detector design have to be considered targeting reduction of the indirect-to-total fraction.

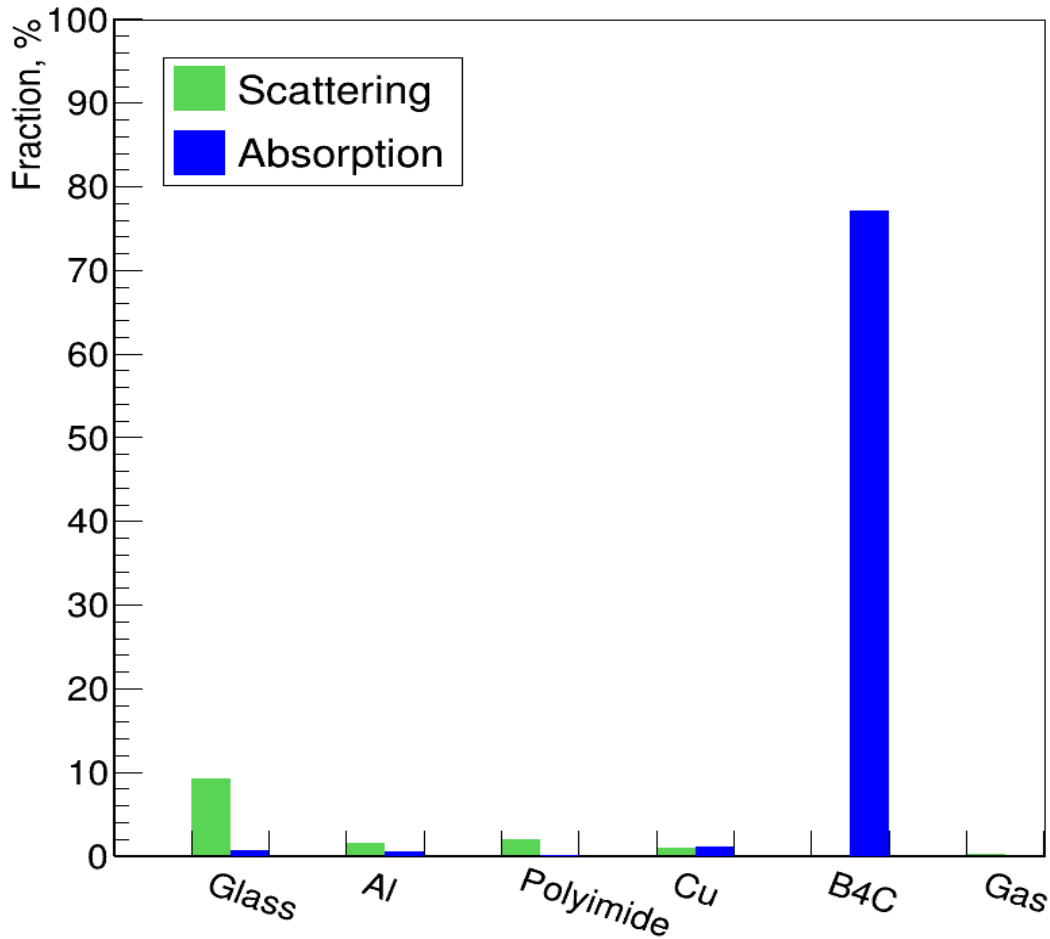

Figure 7. Fractions (in percents) of the number of neutrons in the beam scattered (teal) and absorbed (blue) in the stack of 10 double-gap hybrid RPCs according to ANTS2 simulations.

Since glass is the main contributor to scattering (see figure 7), the amount of this material in the neutron beam path should be reduced as much as possible. Our preliminary results with a small-scale prototype show that it is possible to manufacture RPC anodes with the glass thickness of 0.28 mm. We consider that it should be possible to completely avoid usage of polyimide films by depositing the signal pick-up electrodes directly on the back surfaces of the resistive anodes. This will also allow to change material of the pick-up electrodes to aluminium thus avoiding usage of copper. The amount of aluminium can be significantly reduced by decreasing the thickness of the cathodes to 0.3 mm. Below this value the electrode plates become too flexible which can have negative effect on the uniformity of the gas-gap width.

Modifying the detector model according to the changes described above (0.28 mm glass anodes, 0.3 mm cathodes, no polyimide insulators and pick-up electrodes made of aluminium), we obtain a reduction of the indirect-to-total event ratio by a factor of two: from 0.133 to 0.066. These changes also result in an increase in the total detection efficiency from 60.3% to 62.4% (for λ = 4.73 Å and using 100 keV threshold).

## 3.3 Counting rate capability

As shown in figure 4, the first double-gap RPC of the stack has the highest contribution to the total detection efficiency, and the contributions of the following RPCs decrease exponentially: since the converter layers have the same thickness, all RPCs attenuate the beam by the same factor.

Previously we have reported a counting rate density limit of about 1 kHz/cm$^2$ for a single gap $^{10}$B-RPC [4]: above this value, the RPC counting rate does not scale linearly with the neutron flux. A detector with a stack of RPCs should be able to operate at a higher rate due to distribution of the capture events over several RPCs. In order to maximize the overall counting rate capability of the detector, all individual RPCs should have, ideally, the same contribution to the total detection efficiency (and, hence, the same counting rate). To approach these conditions, an optimization of the individual thicknesses of the $^{10}$B$_4$C converter layers has to be performed.

Two additional aspects have to be taken into account in such optimization. Our simulations indicate that in the case when the contributions from all RPCs to the total detection efficiency are nearly the same, the total efficiency becomes quite small. This is most likely a consequence of the fact that the maximum detection efficiency of a single RPC for thermal neutrons is ≤10% [1]. If the contribution of the last RPC in the stack is equal to that of the other ones, a significant fraction of the neutrons of the beam exit the detector without interaction.

Therefore, the optimization procedure has to find a configuration which not only has a high degree of equality in contributions from individual RPCs, but also results in as high as possible overall detection efficiency of the detector. Since this task requires multi-parameter optimization procedure, we have decided to apply the approach described in section 2.3. The following goodness parameter F is chosen:

$$F = D \cdot E \quad ,$$

where **D** is the overall *direct* detection efficiency of the detector (efficiency which takes into account only those neutrons which did not have prior scattering) and **E** is the equality parameter:

$$E = \frac{\sum_{i=1}^{N} d_i}{N \cdot d_{max}} \quad ,$$

where $d_i$ is the contribution to the total detection efficiency from the RPC with the index $i$, $N$ is the number of RPCs in the stack and $d_{max}$ is the maximum value of $d_i$ in the stack. The equality parameter with the value of unity (maximum possible value) corresponds to the case when all RPCs contribute equally, and, hence, can operate with the same counting rate. For smaller values some of the RPCs operate at a rate lower than optimal. The equality parameter provides the fractional loss in the maximum counting rate compared to the most optimistic scenario of equal contributions. Note that for the prototype configuration with the same thickness of 1.15 µm for all $^{10}B_4C$ layers the equality parameter has a value of 0.367. The minimization procedure described in chapter 2.3 operates with a cost function which returns a negative of the goodness parameter $F$.

The second aspect which has to be considered in the optimization is the practical limitations on the $^{10}B_4C$ layer thicknesses. Firstly, for a hybrid RPC, the $^{10}B_4C$ layers are deposited on both sides of a thin aluminium cathode plate. In order to void deformation of the plate, it is preferable to have the same $^{10}B_4C$ layer thickness on both sides so that the integral of the stress generated by both layers on the plate is null. Therefore, in this study the optimization is performed assuming the same thickness of the converter layer on both sides of the cathode.

Secondly, manufacture of a set of cathode plates with many different converter thicknesses increases the production and quality control time impacting on the cost of the detector. Therefore, the performance of the detector has to be compared for scenarios in which several RPCs of the stack have the same layer thicknesses. In this study we have decided to compare three cases with 3, 5 and 10 different layer thicknesses.

The optimization study is performed using the ANTS2 toolkit benefiting from the infrastructure already created for data analysis and semi-automatic optimization (see section 2.3). On each iteration step a simulation is performed irradiating the detector normally with a pencil beam of $2 \cdot 10^5$ neutrons of 3.656 meV. The detection threshold for the deposited energy is set to 100 keV. The configuration details of the optimization procedure for these three cases are given below followed by a subsection discussing the results.

## a) Five different layer thicknesses

The configuration in which there are five different $^{10}B_4C$ layer thicknesses offers a good compromise between the flexibility in the adjustment and complexity of the detector. The RPC stack is arranged in such a way that consecutive double-gap hybrid RPCs have in pairs the same converter layer thickness (for the RPC indexes from 1 to 10 the thicknesses are: $t_1$ $t_1$ $t_2$ $t_2$ $t_3$ $t_3$ $t_4$ $t_4$ $t_5$ $t_5$).

The optimization procedure started with all layer thicknesses of 1 µm and the initial step of the minimizer of 0.3 µm. The optimization took about 20 minutes on a consumer-grade PC (i7-4790 processor), with the best value of the goodness parameter obtained at the iteration number 87. Figure 8 shows the evolution of the goodness parameter with iteration number, demonstrating a factor of two increase in comparison to the initial value.

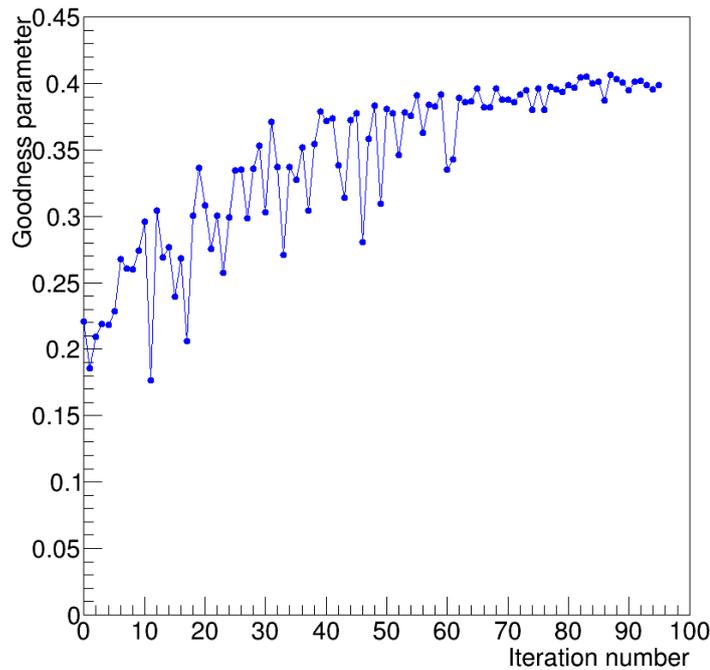

Figure 8. Evolution of the goodness parameter with iteration number of optimization.

The resulting configuration has the layer thicknesses of 0.36, 0.46, 0.72, 1.23 and 2.54 µm. The contributions to the total and the direct (events with no prior scattering) detection efficiencies from the individual double gap RPCs are shown in figure 9. The obtained values of the equality parameter, the overall detection efficiency and the indirect-to-total event ratio are given in Table 2.

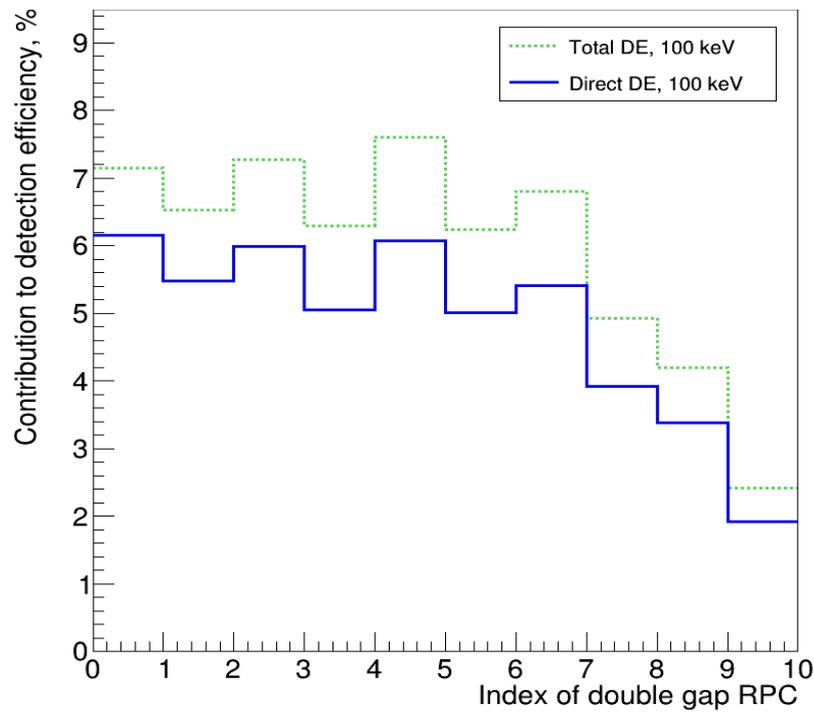

Figure 9. Contributions to the total detection efficiency (dotted green line) and the direct efficiency (straight blue line) from the individual double gap RPCs of the stack for the optimized configuration with five individual thicknesses.

*b) 10 different layer thicknesses*

In this case each double-gap RPC in the stack has individually-adjusted thickness of the converter layers. The optimization started with all thicknesses of 1 μm and the initial step of 0.3 μm. An optimum was found after 173 iterations with the layer thicknesses of 0.37, 0.42, 0.48, 0.58, 0.76, 0.99, 0.93, 0.72, 1.37 and 1.36 μm. The contributions to the total and the direct detection efficiencies from the individual double gap RPCs are shown in figure 10 (left) and the resulting equality parameter, the total detection efficiency and the indirect-to-total event ratio are 0.759, 60.2% and 0.188, respectively. A new round of optimization starting from the obtained layer thicknesses and using 0.3 μm initial step did not result in an improvement in the goodness parameter.

Another optimization run was also performed. This time, considering the results of the previous optimizations, the starting thicknesses of the layers were chosen to be 0.30, 0.40, 0.50, 0.60, 0.70, 0.80, 1.00, 1.40, 2.00 and 2.50 μm. After 53 iterations, an optimum was found with the layer thicknesses of 0.34, 0.39, 0.44, 0.54, 0.60, 0.83, 1.19, 1.25, 2.07 and 3.33 μm. The resulting equality parameter, the total detection efficiency and the indirect/total event ratio are 0.819, 59.7% and 0.187, respectively, and the contributions to the total and the direct detection efficiencies are shown in figure 10 (right). The results are nearly the same as obtained in the first run, and a small increase in the equality parameter is balanced by a small decrease in the detection efficiency.

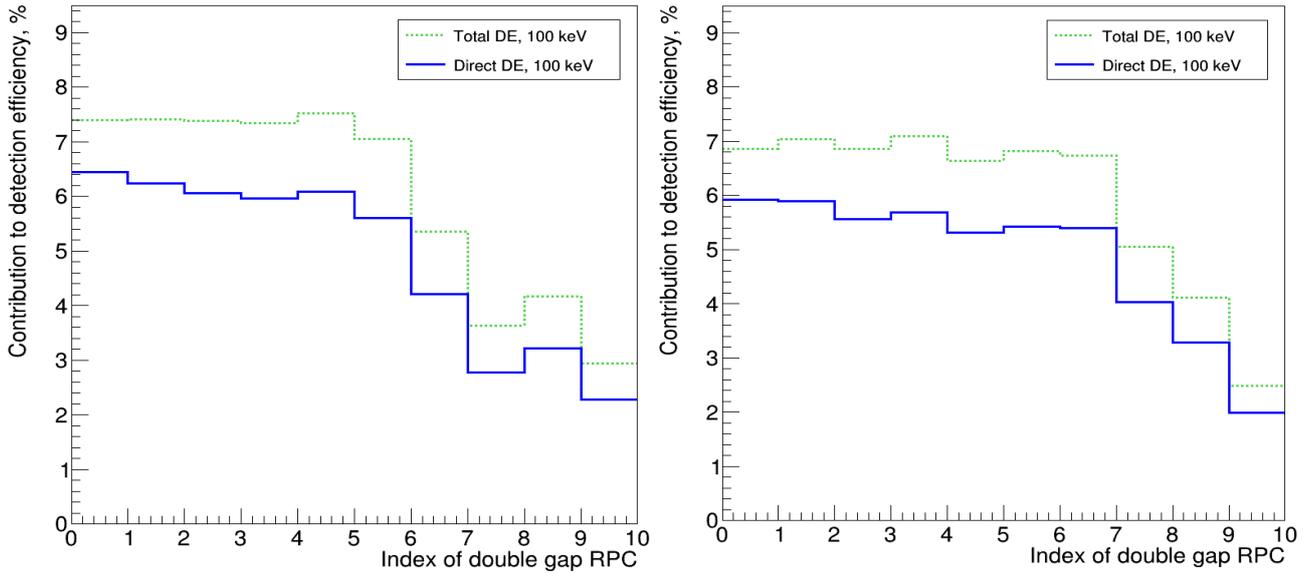

Figure 10. Contributions to the total detection efficiency (dotted green line) and the direct efficiency (straight blue line) from the individual double gap RPCs of the stack for the optimized configuration with ten individual thicknesses. The results are shown for two optimization runs with different initial conditions.

*c) 3 different layer thicknesses*

Based on analysis of the distributions of the layer thicknesses obtained for the previously considered configurations, we have decided to use the same layer thickness for the groups of the first 3, the following 4, and the last 3 double gap RPCs (for the RPC indexes from 1 to 10 the thicknesses are: $t_1$ $t_1$ $t_1$ $t_2$ $t_2$ $t_2$ $t_2$ $t_3$ $t_3$ $t_3$). The starting layer thicknesses were selected to be 0.4, 0.8 and 2.2 µm. After 55 iterations, the optimization resulted in the layer thicknesses of 0.45, 0.8 and 2.01 µm. The resulting equality parameter, the total detection efficiency and the indirect-to-total event ratio are 0.662, 60.7% and 0.175, respectively.

Restarting optimization from these layer thicknesses (initial step is again 0.3 µm) resulted in a different combination of the layer thicknesses: 0.35, 0.53 and 1.43 µm. For this configuration the equality parameter, the total detection efficiency and the indirect-to-total event ratio are 0.759, 57.9% and 0.198, respectively. It seems that there are at least two minima with essentially the same best goodness parameter value: configurations belonging to the first one give higher detection efficiency while configurations from the other one provide better equality parameter.

The contributions to the total and the direct detection efficiencies from the individual double gap RPCs are shown in figure 11 (the results for the first and the second rounds are on the left and right hand sides, respectively).

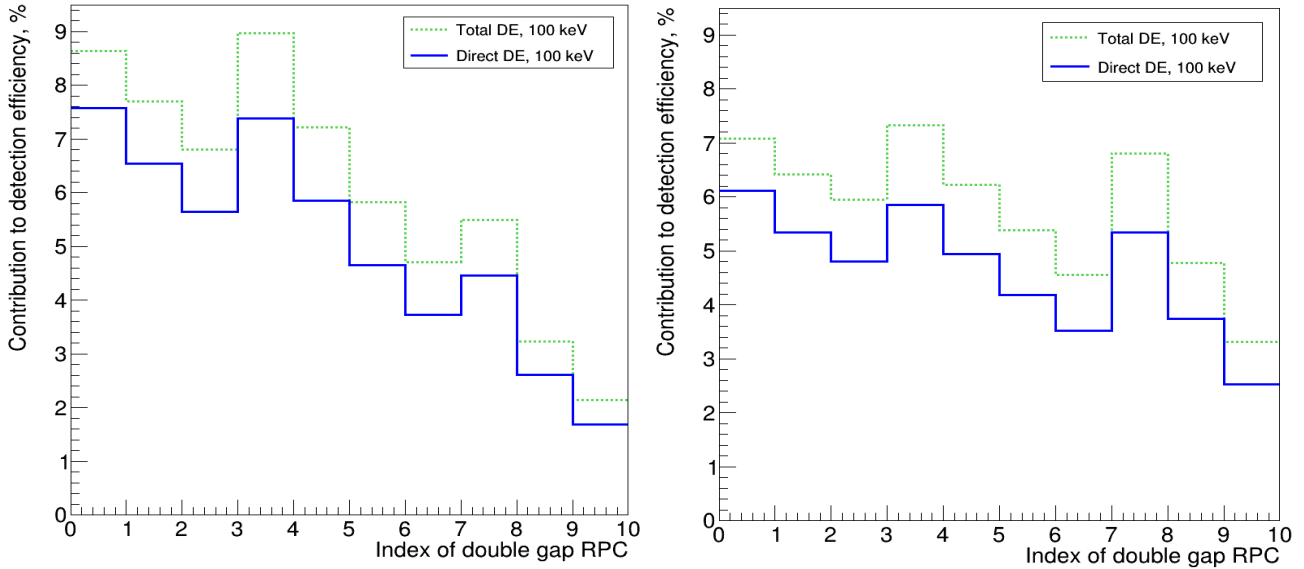

Figure 11. Contributions to the total detection efficiency (dotted green line) and the direct efficiency (straight blue line) from the individual double gap RPCs of the stack for the optimized configuration with three individual thicknesses. The results are shown for two optimization runs with different initial conditions.

### *d) Comparison of the optimization results*

A compilation of the results obtained for the optimized configurations is given in table 2.

| # of individual layer thicknesses | Equality parameter | Direct detection efficiency, % | Total detection efficiency, % | Indirect-to-total ratio |
|---|---|---|---|---|
| 1 (prototype) | 0.367 | 52.3 | 60.3 | 0.133 |
| 3 | 0.662 | 50.1 | 60.7 | 0.175 |
| 3 | 0.759 | 46.4 | 57.9 | 0.198 |
| 5 | 0.786 | 48.4 | 59.5 | 0.186 |
| 10 | 0.759 | 48.9 | 60.2 | 0.188 |
| 10 | 0.819 | 48.5 | 59.7 | 0.187 |

Table 2. Performance parameter values of the detector obtained in optimizations with 3, 5, and 10 different layer thicknesses. The data for the configuration of the first prototype (all layers have the same thickness) are shown for comparison. The statistical uncertainty is ±1 in the last digit.

The results show that it is possible to obtain a factor of two increase in the maximum counting rate compared to the detector prototype with all layers of the same thickness. However, the total detection efficiency becomes lower by several percents, and the indirect-to-total event fraction becomes 1.5 times larger due to the increase of the average path of neutrons inside the detector.

The results also demonstrate that already with three different converter layer thicknesses it is possible to significantly improve the maximum counting rate capability. Increase in the number of different thicknesses from 3 to 5 allows to slightly improve the detection efficiency and the equality parameter. Further increase to 10 individual layers essentially does not improve the results.

# 4   Conclusions

The results of this study demonstrate that a multilayer $^{10}$B-RPC thermal neutron detector with a stack of 10 double gap hybrid RPCs can operate in linear regime with rate densities of ≈8 times higher in comparison to a detector with only one double-gap RPC. This can be achieved by optimizing the converter layer thicknesses of individual RPCs. More than a factor of two increase in the counting rate capability without reduction in the total detection efficiency is obtained in comparison to the configuration of the recently tested detector prototype with all converter layers of the same thickness.

The fraction of detected events from indirect neutrons (those which have at least one scatter before detection) increases from 13% for the prototype configuration to ≈19% for the configurations optimized for the counting rate due to increase in the average path of neutrons inside the RPC stack. However, we show that a factor of two reduction in the indirect-to-total ratio should be possible to achieve for the prototype configuration by decreasing the thickness of the glass anodes and aluminium cathodes to 0.28 mm and 0.3 mm, respectively, and by avoiding the use of polyimide insulators.

# 5   Acknowledgments

This work was supported by the European Union's Horizon 2020 research and innovation programme under grant agreement No 654000.